\documentstyle[12pt]{article}
\title{Some eigenstates for a model associated with solutions
       of tetrahedron equation.\\
       III.~Tetrahedral Zamolodchikov algebras
       and perturbed strings}
\author{I.G.~Korepanov\\
\footnotesize $\matrix{ \cr
       \hbox{Chelyabinsk University of Technology}\cr
       \hbox{76 Lenin av., Chelyabinsk 454080, Russia}
       }$}
\date{March 1997}
\def\be{\begin{equation}}
\def\ee{\end{equation}}
\makeatletter
\long\def\@makecaption#1#2{\vskip 10\p@ \hbox to\hsize{\hfil#1\hfil}}
\makeatother
\begin{document}
\maketitle

\begin{abstract}
This paper continues the series begun with works solv-int/9701016
and solv-int/9702004. Here we show how to construct
eigenstates for a model based on tetrahedron equation
using the tetrahedral Zamolodchikov algebras.
This yields, in particular, new eigenstates for the model
on infinite lattice---`perturbed', or `broken', strings.
\end{abstract}

\section*{Introduction}

In this work we continue the study, begun in~\cite{I,II}, of eigenstates
of the model based on solutions to the tetrahedron equation.
The solutions can be found in paper~\cite{k2}. The method we will be using
is based upon the trigonometrical tetrahedral Zamolodchikov algebras
described in~\cite{k-trig}.

The idea is that sometimes we can control the evolution under the action
of transfer matrix powers for the states arising from a (kagome)
lattice of five-legged `$R$-operators', with given boundary conditions.
Here we continue to use
the notations of \cite{k-alg-geo,k-trig,k2}
where a four-legged object is called~`$L$',
a five-legged one is called~`$R$', and a six-legged one is called~`$S$'.
The object `$L$' is a usual $1+1$-dimensional $L$-operator obeying
the free-fermion condition. $L$'s are not used directly in this paper.
The object `$R$' is, roughly speaking, some special pair of $L$'s.
The fifth leg of $R$ serves to bear an additional superscript
taking values 0 and 1 and intended to mark two elements within the pair.

Some special pair of $R$'s enters in the defining relation
of a {\em tetrahedral Zamolodchikov algebra}, namely
\be
R_{12}^a \tilde R_{13}^b \skew5 \tilde{\tilde R}\vphantom R_{23}^c =
 \sum_{d,e,f} S_{def}^{abc}
 \skew5 \tilde{\tilde R}\vphantom R_{23}^f \tilde R_{13}^e R_{12}^d.
\label{tza}
\ee
This is illustrated by Figure~\ref{figIII-1}.
\begin{figure}[ht]
\begin{center}
\unitlength=1.00mm
\linethickness{0.4pt}
\begin{picture}(134.00,63.00)
\put(77.00,8.00){\line(4,1){56.00}}
\put(86.00,33.00){\line(3,-1){48.00}}
\put(80.00,0.00){\line(1,2){20.00}}
\put(85.00,10.00){\line(1,4){13.00}}
\put(125.00,20.00){\line(-1,1){38.00}}
\put(95.00,30.00){\line(0,1){33.00}}
\put(85.00,10.00){\circle*{2.00}}
\put(95.00,30.00){\circle*{2.00}}
\put(125.00,20.00){\circle*{2.00}}
\put(95.00,50.00){\circle*{2.00}}
\put(2.00,18.00){\line(4,1){56.00}}
\put(1.00,23.00){\line(3,-1){48.00}}
\put(35.00,0.00){\line(1,2){20.00}}
\put(50.00,30.00){\line(1,4){6.00}}
\put(10.00,20.00){\line(-1,1){10.00}}
\put(40.00,10.00){\line(0,1){35.00}}
\put(0.00,31.00){\makebox(0,0)[lb]{$a$}}
\put(39.00,45.00){\makebox(0,0)[rb]{$b$}}
\put(55.00,54.00){\makebox(0,0)[rb]{$c$}}
\put(87.00,57.00){\makebox(0,0)[rt]{$a$}}
\put(94.00,63.00){\makebox(0,0)[rc]{$b$}}
\put(99.00,62.00){\makebox(0,0)[lc]{$c$}}
\put(0.00,23.00){\makebox(0,0)[rc]{1}}
\put(2.00,17.00){\makebox(0,0)[ct]{2}}
\put(34.00,1.00){\makebox(0,0)[rb]{3}}
\put(85.00,33.00){\makebox(0,0)[rc]{1}}
\put(77.00,9.00){\makebox(0,0)[cb]{2}}
\put(79.00,0.00){\makebox(0,0)[rb]{3}}
\put(10.00,20.00){\circle*{2.00}}
\put(40.00,10.00){\circle*{2.00}}
\put(50.00,30.00){\circle*{2.00}}
\put(66.00,22.00){\makebox(0,0)[cc]{\Large$=$}}
\put(109.00,37.00){\makebox(0,0)[lb]{$d$}}
\put(96.00,42.00){\makebox(0,0)[lc]{$e$}}
\put(87.00,22.00){\makebox(0,0)[rb]{$f$}}
\put(10.00,22.00){\makebox(0,0)[lb]{$R$}}
\put(40.00,8.00){\makebox(0,0)[lt]{$\tilde R$}}
\put(50.00,29.00){\makebox(0,0)[lt]{$\skew4 \tilde{\tilde R}$}}
\put(97.00,51.00){\makebox(0,0)[lc]{$S$}}
\put(85.00,9.00){\makebox(0,0)[lt]{$\skew4 \tilde{\tilde R}$}}
\put(95.00,28.00){\makebox(0,0)[lt]{$\tilde R$}}
\put(125.00,22.00){\makebox(0,0)[lb]{$R$}}
\end{picture}
\end{center}
\caption{}
\label{figIII-1}
\end{figure}

Suppose we have fixed some boundary conditions in the tensor product
of spaces denoted 1, 2 and~3 (which means, most generally, that we have
taken the trace of a product of each side of~(\ref{tza}) and some
linear operator acting in the mentioned tensor product).
This yields, in the l.h.s.\ of~(\ref{tza}), some vector in the tensor
product of spaces corresponding to indices $a$, $b$ and $c$, and in the
r.h.s.\ of~(\ref{tza})---the result of $S$-operator action upon
a similar vector. For different boundary conditions, this provides
enough (consistent) relations for $S$-operator to be determined
uniquely.

We will look at this, however, from another point of view, using
boundary conditions for $R$'s as a means to define vectors in the
space where $S$'s act. Of course, we will
take, instead of just three $R$-operators,
a large lattice made up of them, whose fragment is depicted
in Figure~\ref{figIII-2},
\begin{figure}[ht]
\begin{center}
\unitlength=1.00mm
\linethickness{0.4pt}
\begin{picture}(120.00,50.00)
\put(25.00,0.00){\line(1,1){50.00}}
\put(45.00,0.00){\line(1,1){50.00}}
\put(65.00,0.00){\line(1,1){50.00}}
\put(35.00,15.00){\line(1,0){50.00}}
\put(55.00,35.00){\line(1,0){50.00}}
\put(20.00,0.00){\line(2,1){100.00}}
\put(40.00,0.00){\line(2,1){60.00}}
\put(60.00,0.00){\line(2,1){20.00}}
\put(40.00,20.00){\line(2,1){60.00}}
\put(60.00,40.00){\line(2,1){20.00}}
\put(30.00,5.00){\line(0,1){3.00}}
\put(50.00,5.00){\line(0,1){3.00}}
\put(70.00,5.00){\line(0,1){3.00}}
\put(40.00,15.00){\line(0,1){3.00}}
\put(50.00,15.00){\line(0,1){3.00}}
\put(60.00,15.00){\line(0,1){3.00}}
\put(70.00,15.00){\line(0,1){3.00}}
\put(80.00,15.00){\line(0,1){3.00}}
\put(90.00,25.00){\line(0,1){3.00}}
\put(70.00,25.00){\line(0,1){3.00}}
\put(50.00,25.00){\line(0,1){3.00}}
\put(60.00,35.00){\line(0,1){3.00}}
\put(70.00,35.00){\line(0,1){3.00}}
\put(80.00,35.00){\line(0,1){3.00}}
\put(90.00,35.00){\line(0,1){3.00}}
\put(100.00,35.00){\line(0,1){3.00}}
\put(70.00,45.00){\line(0,1){3.00}}
\put(90.00,45.00){\line(0,1){3.00}}
\put(110.00,45.00){\line(0,1){3.00}}
\end{picture}
\end{center}
\caption{}
\label{figIII-2}
\end{figure}
and apply a layer of $S$-operators---a
hedgehog transfer matrix---to it.
We will see that sometimes, for simple boundary
conditions, this can be a reasonable way of describing the vectors
on which the transfer matrix acts, as well as the results
of such action.

\smallskip

The $R$-operators we will be dealing with in this paper are the simplest
possible---trigonometrical---ones. In this connection, let us refer
to Theorem~2.3 of the work~\cite{k2} wherefrom it follows that to
an $S$-matrix corresponds a {\em two-parameter\/} family of
triples of $R$-operators. If we restrict ourselves to only trigonometrical
$R$-matrices from~\cite{k-trig}, then there remains a {\em one-parameter\/}
family of those. So, below it is implied that we are constructing
one-parameter families of states for a given transfer matrix.

\section{Two kinds of strings on a finite lattice}
\label{sec-finite}

\subsection{Eigenstates with eigenvalue~1 yielded by the lattice
with a given ``polarization''}
\label{topological}

For a finite lattice on a torus, the ``periodic'' boundary
conditions seem, at first sight, to be already fixed.
However, trigonometrical $R$-operators of work~\cite{k-trig} conserve
the ``number of particles'', sometimes called also ``polarization''
(because they are very much like the usual 6-vertex model
$L$-operators), and this provides more possibilities.
Namely, the reader can easily verify that the following construction
yields some states that are transformed into themselves by the hedgehog
transfer matrix.

Let us declare some edges of the kagome lattice (Figure~\ref{figIII-2})
`black' and the others `white' in such a way that the number of black
lines is conserved at each vertex (the incoming edges being situated
below and to the left of the vertex, and the outgoing edges---above
and to the right). It can be said that such a configuration of black
edges---we will call it {\em permitted\/} configuration---forms
a cycle belonging to some homology class of the torus.
Let us say that vectors $\pmatrix{0\cr 1}$ correspond to white
edges, and vectors $\pmatrix{1\cr 0}$---to black edges.
This selects some matrix element for each $R$-operator,
but as there are really two operators numbered by the upper index,
this selects a pair of numbers forming a vector in the
two-dimensional space corresponding to a {\em vertical\/} edge
in Figure~\ref{figIII-2}. The tensor product of such vectors lies
in the space where the transfer matrix acts.
Now let us take a sum of those vectors
over all black edges configurations belonging
to the same homological class.
It is quite straightforward to see that the transfer matrix transforms
this sum into exactly the same sum.

\subsection{How the moving strings arise from the lattice
of $R$-operators}

A slight modification of the above construction yields the moving
strings from work~\cite{I}. Namely, fix arbitrarily some straight
lines of the kagome lattice of Figure~\ref{figIII-2} and paint black
all the edges belonging to them. Then those lines will move under
the action of transfer matrix as described in~\cite{I}.

The only nontrivial point here is with {\em oblique\/} lines.
To make this clear, let us draw some more pictures.
First, let us interchange l.h.s.\ and r.h.s. in Figure~\ref{figIII-1}
and redraw it like the following formula:
$$
S \sum \hbox{configurations of } \matrix{
\unitlength=1mm
\linethickness{0.4pt}
\begin{picture}(14.00,14.00)
\put(0.00,0.00){\line(1,1){14.00}}
\put(4.00,0.00){\line(0,1){14.00}}
\put(0.00,10.00){\line(1,0){14.00}}
\end{picture}
} = \sum \hbox{configurations of } \matrix{
\unitlength=1mm
\linethickness{0.4pt}
\begin{picture}(14.00,14.00)
\put(0.00,0.00){\line(1,1){14.00}}
\put(0.00,4.00){\line(1,0){14.00}}
\put(10.00,0.00){\line(0,1){14.00}}
\end{picture}
},
$$
where ``configurations'' means ``vectors corresponding to
permitted configurations of three black
edges within a triangle with given `boundary condition' for six
external edges''.
In these terms, Figure~\ref{figIII-1} itself says only that
$$
S \left( \matrix{
\unitlength=1mm
\linethickness{0.4pt}
\begin{picture}(14.00,14.00)
\put(0.00,0.00){\line(1,1){14.00}}
\end{picture}
} + \matrix{
\unitlength=1mm
\linethickness{0.4pt}
\begin{picture}(14.00,14.00)
\put(0.00,0.00){\line(1,1){4.00}}
\put(4.00,4.00){\line(0,1){6.00}}
\put(4.00,10.00){\line(1,0){6.00}}
\put(10.00,10.00){\line(1,1){4.00}}
\end{picture}
}\right) = \matrix{
\unitlength=1mm
\linethickness{0.4pt}
\begin{picture}(14.00,14.00)
\put(0.00,0.00){\line(1,1){14.00}}
\end{picture}
} + \matrix{
\unitlength=1mm
\linethickness{0.4pt}
\begin{picture}(14.00,14.00)
\put(0.00,0.00){\line(1,1){4.00}}
\put(4.00,4.00){\line(1,0){6.00}}
\put(10.00,4.00){\line(0,1){6.00}}
\put(10.00,10.00){\line(1,1){4.00}}
\end{picture}
}
$$
(here only the black edges are depicted), and not that
\be
S \matrix{
\unitlength=1mm
\linethickness{0.4pt}
\begin{picture}(14.00,14.00)
\put(0.00,0.00){\line(1,1){14.00}}
\end{picture}
} = \matrix{
\unitlength=1mm
\linethickness{0.4pt}
\begin{picture}(14.00,14.00)
\put(0.00,0.00){\line(1,1){14.00}}
\end{picture}
}.
\label{oblique}
\ee
However, (\ref{oblique}) is proved by direct calculation involving
the explicit expressions for matrix elements of $R$-operators.

\section{`Broken' infinite strings}
\label{sec-lom}

Consider now the infinite kagome lattice.
Let the edges painted black be placed in such manner that they form
two horizontal rays, one going to the right and one going to
(or rather coming from) the left, as in Figure~\ref{figIII-3},
\begin{figure}[ht]
\begin{center}
\unitlength=1.00mm
\linethickness{0.4pt}
\begin{picture}(120.00,40.00)
\put(0.00,5.00){\line(1,0){50.00}}
\put(70.00,35.00){\line(1,0){50.00}}
\multiput(52.00,8.00)(2.00,3.00){9}{\makebox(0,0)[cc]{.}}
\put(51.00,4.00){\makebox(0,0)[lt]{$A$}}
\put(69.00,36.00){\makebox(0,0)[rb]{$B$}}
\end{picture}
\end{center}
\caption{}
\label{figIII-3}
\end{figure}
and those rays be connected by some path going along the lattice
edges and {\em permitted\/} in the sense of Subsection~\ref{topological}.
Consider the vector---the formal infinite tensor product---corresponding
to such black edge configuration, and take a sum over {\em all\/}
the permitted paths linking the two rays (in particular, the ends
$A$ and $B$ of the rays can move anywhere to the left and/or to
the right).

This gives some generalization of the straight strings of paper~\cite{I}
or, at all events, of one such string---we leave the careful
consideration of the case of several strings for a separate work.
As in \cite{I}, formal
eigenvectors can be built out of translations of such string.

\section{Discussion}
\label{III-discussion}

\subsection{Tetrahedral Zamolodchikov algebras and vacuum vectors}

Tetrahedral Zamolodchikov algebras are not, of course, exactly the
same thing that the $S$-operator's vacuum vectors, but they are
intimately connected. To be exact, in general case one can obtain
a {\em modified\/} Zamolodchikov algebra from vacuum vectors as
follows. Let it be known that for some $S$-operator, taken
e.g.\ from the work~\cite{mss},
\be
S(X\otimes Y\otimes Z)=U\otimes V\otimes W,
\label{III_*}
\ee
and let the tetrahedron equation hold (where the dots stand for
the parameters on which the $S$'s depend, each of four $S$'s
having its own parameters):
\be
S_{123}(\ldots)S_{145}(\ldots)S_{246}(\ldots)S_{356}(\ldots)=
S_{356}(\ldots)S_{246}(\ldots)S_{145}(\ldots)S_{123}(\ldots)
\label{III_**}
\ee
In contrast with the paper~\cite{II}, here each of the six spaces
in whose tensor product the $S$-operators act is marked by a
{\em single\/} number. In formula~(\ref{III_*}) those
numbers---superscripts at $S=S_{123}(\ldots)$---are dropped.
We will, if necessary, attach those superscripts at vectors as well,
to mark the number of space where a vector belong, so that
(\ref{III_*}) will become
$$
S_{123}(\ldots)(X_1 \otimes Y_2 \otimes Z_3)=
 U_1 \otimes V_2 \otimes W_3.
$$

Now let us apply both sides of (\ref{III_*}) to the product
$(X_1 \otimes Y_2 \otimes Z_3)$. We will get an equation
like~(\ref{tza}), but with six, generally speaking, {\em different\/}
$R$'s:
$$
SR_{45}(\ldots)R_{46}(\ldots)R_{56}(\ldots)=
R'_{56}(\ldots)R'_{46}(\ldots)R'_{45}(\ldots),
$$
where e.g.\
$$
R_{45}(\ldots)=S_{145}(\ldots) X_1, \qquad
R'_{45}(\ldots)=S_{145}(\ldots) U_1
$$
etc. It is clear that a ``usual'' Zamolodchikov algebra will
appear if
$$
X=U, \qquad Y=V, \qquad Z=W.
$$

Probably, the obvious connections between the
tetrahedral Zamolodchikov algebras and vacuum vectors, described here,
are worth further investigation.

\subsection{General conclusions}

The papers~\cite{I,II} and the present one show that
\begin{itemize}
\item
even the model corresponding to the simplest solutions of tetrahedron
equation possesses a large variety of eigenstates which are probably
not easy to classify,
\item
eigenvalues seem sometimes rather trivial---maybe, it is due to
relation $S^2={\bf 1}$, see~\cite{k2}---but probably they will be
more interesting for the models from \cite{h,mss},
\item
some states can be introduced that are countable sums of
formal tensor products
throughout the infinite lattice, but it is unclear what to do
for a finite lattice,
\item
there probably does not exists---at least, it was not
discovered in papers~\cite{h,mss}---a complete analog
of the 6-vertex model in $1+1$ dimensions with its
``conservation of particle number'', but something can
be built even upon the fact that that number is conserved ``sometimes'',
\item
there exists a huge amount of symmetries multiplying eigenvalues
by constants (roots of unity for a finite lattice) and
unknown for the $1+1$-dimensional models, and
\item
eigenstates can be constructed with making no use of {\em invertibility\/}
of tetrahedron equation solutions---so probably it makes sense to search
for non-invertible ones.
\end{itemize}

\end{document}